\documentclass[journal]{journal}

\usepackage{multicol,lipsum}
	\usepackage[pdftex]{graphicx}
\usepackage[cmex10]{amsmath}

\usepackage{mathtools}

\usepackage{float}

\usepackage[export]{adjustbox}

\usepackage{fixltx2e}
\usepackage{ulem}

\usepackage{xcolor}

\usepackage{nameref}

\begin{document}
%
\title{Direct Participation of Dynamic Virtual Power Plants in Secondary Frequency Control}
%
%
%

\author{{M. Ebrahim ADABI,
        Bogdan MARINESCU}
        }
        
\thanks{M. Ebrahim ADABI and Bogdan MARINESCU are with Ecole Centrale Nantes -LS2N (Laboratoire des sciences du numérique de Nantes), 1 Rue de la Noë, 44000 Nantes Cedex 3, France, Email: ebrahim.adabi@ec-nantes.fr, Bogdan.Marinescu@ec-nantes.fr}
\thanks{This work is part of the H2020 European project POSYTYF (https://posytyf-h2020.eu/).}

%
%

\markboth{Journal of \LaTeX\ Class Files,~Vol.~6, No.~1, January~2007}%
{Shell \MakeLowercase{\textit{et al.}}: Bare Demo of IEEEtran.cls for Journals}
%



\maketitle
\thispagestyle{empty}
\begin{abstract}
This paper proposes a novel control strategy in which Renewable Energy Sources (RES) considered in a new Dynamic Virtual Power Plant (DVPP) concept directly participate to Secondary Frequency Control (SFC). This allows full participation of these generators to SFC, i.e., in the same manner as classic synchronous generators by fulfilling identical specifications from both control and contractual points of view. An internal real-time redispatch has been proposed to account in DVPP in order to determine the amount of active power injection by each RES unit for provision of frequency support in secondary level. The whole control scheme is designed to take into account both rapid and slow dynamics of modern power systems which contain both classic synchronous generators and rapid power electronics for renawable energy sources in which DVPP is supposed to be inserted. The performance of secondary frequency control strategy has been validated through simulation studies on a two-area benchmark with mixed wind power plants and classic synchronous generators. This work is part of the H2020 POSYTYF project (https://posytyf-h2020.eu/).
\end{abstract}


%

\section{Introduction}\label{sectionintroduction}
High penetration of Renewable Energy Sources (RES) in modern power grids is of critical importance for the transformation of the global energy system \cite{Wind Penetration1}, \cite{Wind Penetration2}, \cite{Wind Penetration3}, \cite{Wind Penetration4}. However, stability and participation to ancillary services issues related to RES is a significant challenge that should be taken in to account. Indeed, the RES grid integration faces major limitations when high RES penetration is expected \cite{Intro 1}. A solution to overcome this is to group several RES into a systemic object called Dynamic Virtual Power Plant (DVPP) \cite{DVPP_WPLs}. DVPP is a way to aggregate RES sources to form a portfolio of dispatchable/non-dispatchable RES able to optimally internally redispatch resources in case of meteorological and system variations in order to provide sufficient flexibility, reliable power output and grid services.

Due to the massive integration of RES, it is essential that they could contribute to provision of ancillary services such as frequency support. In the previous publications which have been studied in the review papers of \cite{IndirectSFC1}, and \cite{IndirectSFC2}, RES \textit{indirectly} participate in provision of ancillary service. In such indirect participation, RES partially contribute to secondary frequency control through decreasing frequency deviation \cite{IndirectSFC3},\cite{IndirectSFC4},\cite{IndirectSFC5},\cite{IndirectSFC6}.

In this paper the main goal is the integration of DVPP to the existing secondary control frame in order to provide  secondary frequency support in the same manner as classic synchronous generators such as thermal, gas, and hydro. To do so, \textit{direct participation} of DVPP to SFC is taken into account on equal basis (obligations, operation and remuneration) with classic synchronous generators. 

The proposed control approach lays on an \textit{innovatory modeling}. Indeed, classic hypothesis for separation of voltage and frequency dynamics used till no longer hold for RES connected to grid by power electronics as they have both rapid voltage and frequency dynamics. The model introduced in \cite{MarIJC} is used here to combine these dynamics into a simple mathematical object used both for the control and simulation validations. 

Moreover, in order to ensure full participation of DVPP in provision of secondary frequency support, an \textit{internal redispatch} strategy has been considered inside the DVPP, which fulfils the constant amount  of injected active power by DVPP, in order to reduce the need for electrochemical energy storage.

The paper is structured as follows: Section 2 presents the problem formulation and framework of the approach. Section 3 presents the new control for direct participation of DVPP to SFC. Simulation results are presented in Section 4 while Section 5 is devoted to conclusions. 

\section{Problem formulation and framework of the approach}

\subsection{Classic SFC}

The Secondary Frequency Control  (SFC), also called Automatic Generation Control (AGC) or Load Frequency Control (LFC) is a high-level control with two main goals: the first one is to maintain frequency into a desirable range and second one is to control the power exchange programmes between  different control areas \cite{SFC1}.
 More specifically, the secondary regulation service for the frequency should provide following specifications:
 \begin{itemize}
 \item Frequency deviation of each SFC area i should go to zero: $\Delta f_{i}\stackrel{t\rightarrow\infty}{\rightarrow}0,$
 
 \item The deviation of each tie-lines power exchange between areas should go to zero: $\Delta P_{tie}\stackrel{t\rightarrow\infty}{\rightarrow}0$.
  \end{itemize}
 
\begin{figure}[!h]
    \centering
    \includegraphics[scale=0.9]{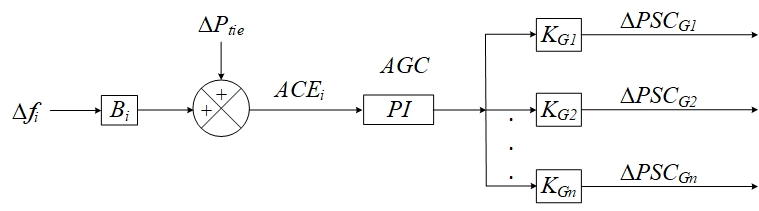}
    \caption{Control block diagram for automatic generation control (AGC)}
    \label{figAGC}
\end{figure}

  These two objectives are usually gathered into a common control signal called Area Control Error (ACE) as in Fig. \ref{figAGC} which shows the general SFC control block diagram. ACE is achieved from the combination of frequency deviation of the area ($\Delta f_{i}$) and the tie-line power between areas ($\Delta P_{tie}$) (\ref{ACE}). $K_{G1}$, $K_{G2}$, and ,$K_{Gn}$, are the participation factors which determine the amount of active power with which each generator participates to SFC ($\Delta PSC_{Gn}$). 
  
    \begin{equation}\label{ACE}
	ACE_i=\Delta P_{tie}+B_i\Delta f_{i}
\end{equation}

\subsection{Indirect Participation of DVPP to classic SFC}

\begin{figure}[!h]
    \centering
    \includegraphics[scale=0.8]{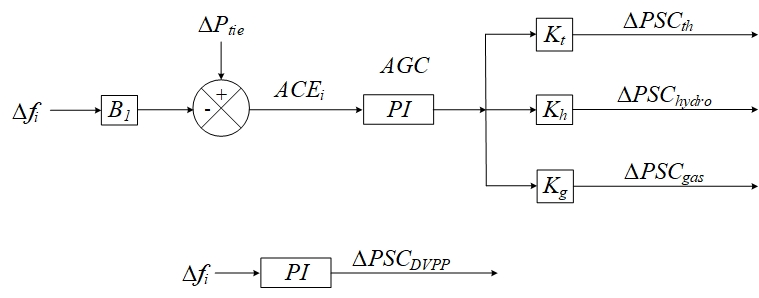}
    \caption{Indirect participation of DVPPs in SFC}
    \label{figInSFC}
\end{figure}

For all up to date SFC control implementations (see, e.g., reviews  \cite{IndirectSFC1}, \cite{IndirectSFC2} which are the two most up to date review papers about the participation of renewable energy sources (especially wind power plants) to frequency control, renewable energy sources \textit{indirectly} contribute to SFC. Fig. \ref{figInSFC}, shows the block diagram for indirect participation of DVPP to SFC as proposed in \cite{IndirectSFC3}, \cite{IndirectSFC4}, \cite{IndirectSFC5}, \cite{IndirectSFC6}. 
Indeed, DVPP only contributes to reduce the frequency deviation of the zone by its independent PI control, while the classic synchronous generators (thermal, hydro, and gas) directly contribute to SFC with participation factors ($K_{t}$,$K_{h}$,$K_{g}$) to correct ACE. 

The RES does not participate in the SFC in the regulatory sense (of contractual obligations and availability of the service). Also, from the technical point of view, their contributions are not quantified; RES ensures some positive impact on frequency deviations when they can (availability of natural resources and unilateral decision to participate) and no contribution to areas power exchange regulation.

To enhance participation of the DVPP to frequency ancillary services, a method for \textit{direct} participation to SFC is proposed in this paper to overcome above mentioned limitations and to allow DVPP to behave like a large classic generator. 

\subsection{New model for simulation and control} \label{SectionControlModels}

In all approaches for RES contribution to SFC, a model based exclusively on swing equations of generators was used (see, e.g.,  \cite{IndirectSFC2}). This is based on the classic hypothesis of separation of voltage and frequency dynamics. This separation is well known and accepted in classical power system where the classic synchronous generators are the main generation units \cite{Kundur}. 

However, due to the massive integration of power electronics converter based  generation unit in power system, the dynamics of voltage and frequency became very fast and in the same range. Therefore the separation of voltage and frequency is not valid anymore for power systems with high penetration of renewable energy source as the DVPP we consider here. A new model that jointly considers the dynamics of frequency and voltage at the same time was introduced in \cite{MarIJC} and will be used here for both control and simulations purposes.

\begin{figure}[H]
    \centering
    \includegraphics[width=0.3\textwidth,center]{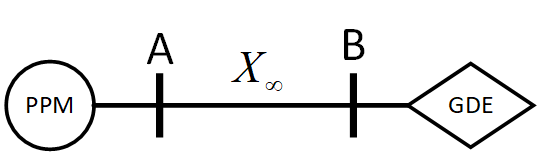}
    \caption{New model for single PPM} 
    \label{fig:control_model_1}   
\end{figure}

Fig. \ref{fig:control_model_1} shows our new concept for modeling in case of one generator (renewable generator or Power Plant Module (PPM)).  It consists of the full model of PPM which has to be controlled, an equivalent AC line of reactance $X_\infty$ and a Grid Dynamic Equivalent (block GDE) which jointly provides the grid voltage and frequency. The line $X_\infty$ accounts for the grid short-circuit power at PPM connection bus A and it is computed in a standard way (see, e.g., \cite{12x}).

In comparison with conventional control models mentioned above, the new model with GDE block has the benefit of capturing the dynamics of voltage and frequency at the same time. The electrical frequency at bus B is not fixed at nominal grid frequency and is determined by GDE block, through the swing equation.
\begin{equation}\label{GDEswing}
	2H\frac{d\omega_f}{dt}=P_G-P_L-D_u\Delta \omega_f
\end{equation}

and the three-phase voltage dynamics are given as:

\begin{equation}\label{GDEvoltage}
\begin{array}{l}
\begin{aligned}
	{V_B}^a=Vsin(\theta_f)\\
	{V_B}^b=Vsin(\theta_f-2\frac{\pi}{3})\\
	{V_B}^c=Vsin(\theta_f+2\frac{\pi}{3}).
\end{aligned}
	\end{array}
\end{equation}

H is the equivalent inertia of the rest of the system (in which the controlled PPM is inserted), $P_G$ is the global active power produced in the rest of the system and $P_L$ corresponds to the global load of the system. $P_m$ is a constant input for the PPM control problem. Dynamics (\ref{GDEswing}) is stabilized by the damping factor $D_u$ and a simple integrator for deviation of $\omega_f$ from the nominal grid frequency which accounts for a secondary frequency control in the rest of the system (GDE). H is computed by classic equivalencing methods (e.g., \cite{Kundur}) used in frequency studies.

Using large scale classic synchronous generator instead of GDE block is also a used solution to model rest of the power system. However, this solution has two drawbacks: first, the voltage and frequency regulations parameters for the physical model of classic synchronous generator would have impact on the resulting dynamics. Second, the order of classic synchronous generator is much higher than the one of the proposed GDE block.

\begin{figure}[H]
    \centering
    \begin{minipage}[b][][b]{0.2\textwidth}
    	\centering
    	\includegraphics[width=1\textwidth,center]{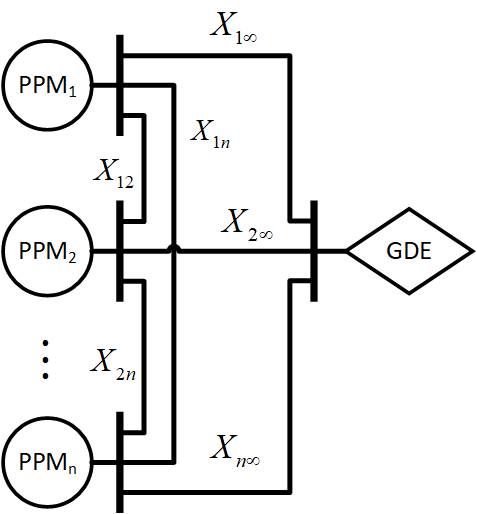}
    \end{minipage}\hfill
    \begin{minipage}[b][][b]{0.24\textwidth}
    	\centering
    	\includegraphics[width=1\textwidth,center]{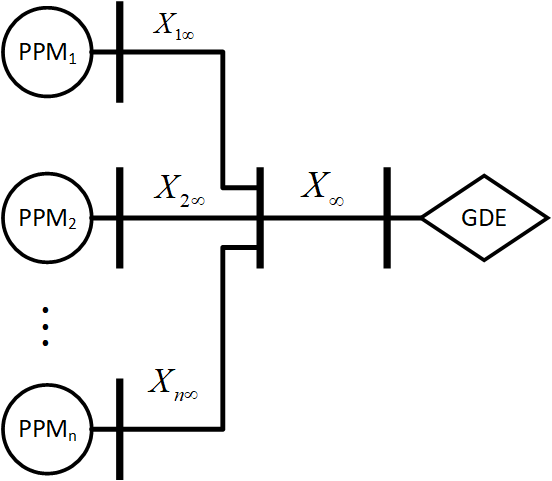}
    \end{minipage}
		\caption{New control model for multiple PPMs}
		\label{controlModelMulti}
\end{figure}

Fig. \ref{controlModelMulti}. shows the structure of proposed control model if several PPMs have to be controlled.

\section{New control for direct participation of the DVPP to the SFC}

\subsection{Direct Participation of the DVPP to the SFC} \label{Dirct Participation}
 
\begin{figure}[!h]
    \centering
    \includegraphics[scale=0.8]{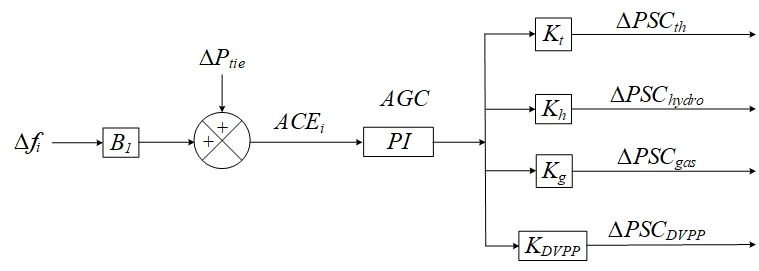}
    \caption{Direct participation of DVPPs in SFC}
    \label{figDISFC}
\end{figure}

Fig. \ref{figDISFC} shows the proposed architecture of control strategy which allows direct SFC  participation of the DVPP. Indeed, the latter is treated in the same manner as classic synchronous generators (thermal, hydro of gas). The DVPP contributes to SFC, with participation factor of $K_{DVPP}$, on equal basis (obligations, operation and remuneration) with classic synronous generators. The sum of DVPP's participation factor ($K_{DVPP}$) and conventional synchronous generator's factor ($K_{t}$,$K_{h}$,$K_{g}$), will be equal to 1.
The new DVPP concept introduced in \cite{DVPP_WPLs} allows one to jointly and optimally manage dispatchable (like, e.g., hydro or solar thermal plants) and non dispatchable (like, e.g., wind and classic solar panels) resources in order to provide predictible and reliable active power output of the DVPP in mid (few minutes) and long time horizons. This allows the DVPP to position on markets but for full SFC integration a faster regulation level is introduced in the next section.

\subsection{Internal DVPP redispatch}\label{Internal Redispatch}
\begin{figure}[!h]
    \centering
    \includegraphics[scale=0.8]{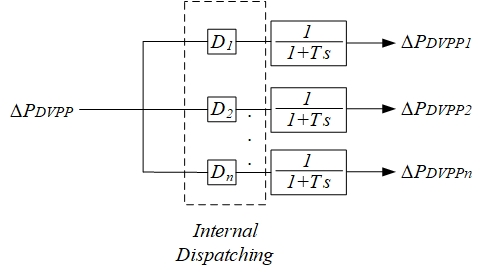}
    \caption{Internal dispatching of the DVPP for participation to the SFC}
    \label{figDispatching}
\end{figure}

The proposed internal redispatch is a supplementary control that allows redistribution of the contribution of each RES of the DVPP in case of rapid changes, intermitance of RES components or other variations. It is slower than the SFC dynamics and faster that the primary frequency regulation in order not to interact with the latter one. 
To avoid a complicated control and need of supplementary measurements, a simple internal regulation is proposed based on time-varying participation factors $D_i$ in Eq. \ref{Redispatch} which are directly evaluated from available active power measure of each RES of the DVPP. If one of the RES is limiting its output or shuts down, coefficients $D_i$ are automatically adjusted. Their sum is always 1.
In order to reach the aforementioned dynamics, a filter with time constant T=4s is implemented as shown in Fig. \ref{figDispatching}.

 Fig. \ref{figDispatching} shows the algorithm for internal dispatching of the DVPP units to contribute to SFC.  $P_{1}$,$P_{2}$,..,$P_{n}$ are the active power generated by each DVPP unit.
   
  \begin{equation}\label{Redispatch}
      \begin{array}{cc}
  D_{1} &= \frac{P_{1}}{P_{1} + P_{2} + ...+P_{n}} \\
D_{2} &= \frac{P_{2}}{P_{1} + P_{2} + ...+P_{n}} \\
\vdots&\\
D_{n} &= \frac{P_{n}}{P_{1} + P_{2} + ...+P_{n}}
      \end{array}
  \end{equation}

\subsection{Simulation benchmark and control strategy for direct participation of the DVPP to the SFC}
\begin{figure}[!h]
    \centering
    \includegraphics[scale=0.4]{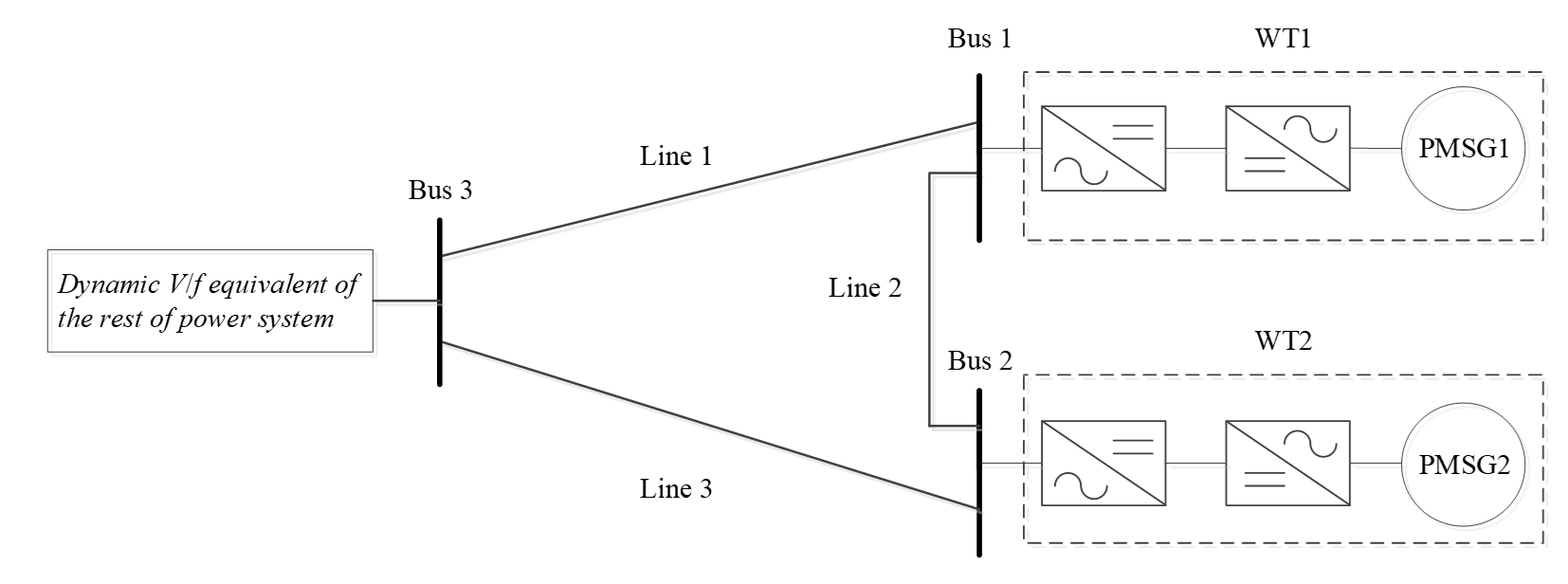}
    \caption{The Benchmark to test the control strategies}
    \label{figBenchmark}
\end{figure}

Fig. \ref{figBenchmark}, shows the benchmark that has been implemented in Matlab/Simulink in order to test the direct participation of the DVPP to SFC. The three bus benchmark is composed of two wind turbines (WT1 and WT2) connected to bus 1 and bus 2 which plays the role of DVPP (n=2), and  a grid dynamic equivalent connected to the bus 3 which provides the grid frequency and also represents the rest of power system (as discussed in Section \ref{SectionControlModels}).
\\

\begin{figure}[!h]
    \centering
    \includegraphics[scale=0.45]{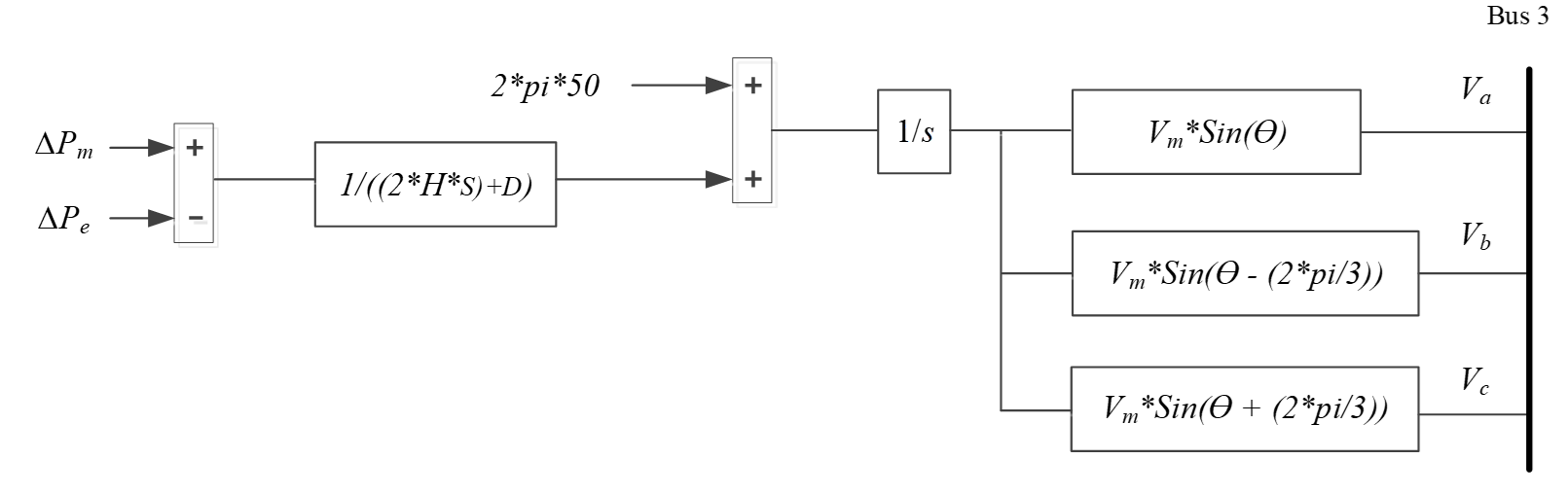}
    \caption{Block diagram for dynamic v/f equivalent of the rest of power}
    \label{figDynamic_vf}
\end{figure}

\begin{figure}[H]	
\centering
\resizebox{0.7\textwidth}{!}
{\includegraphics{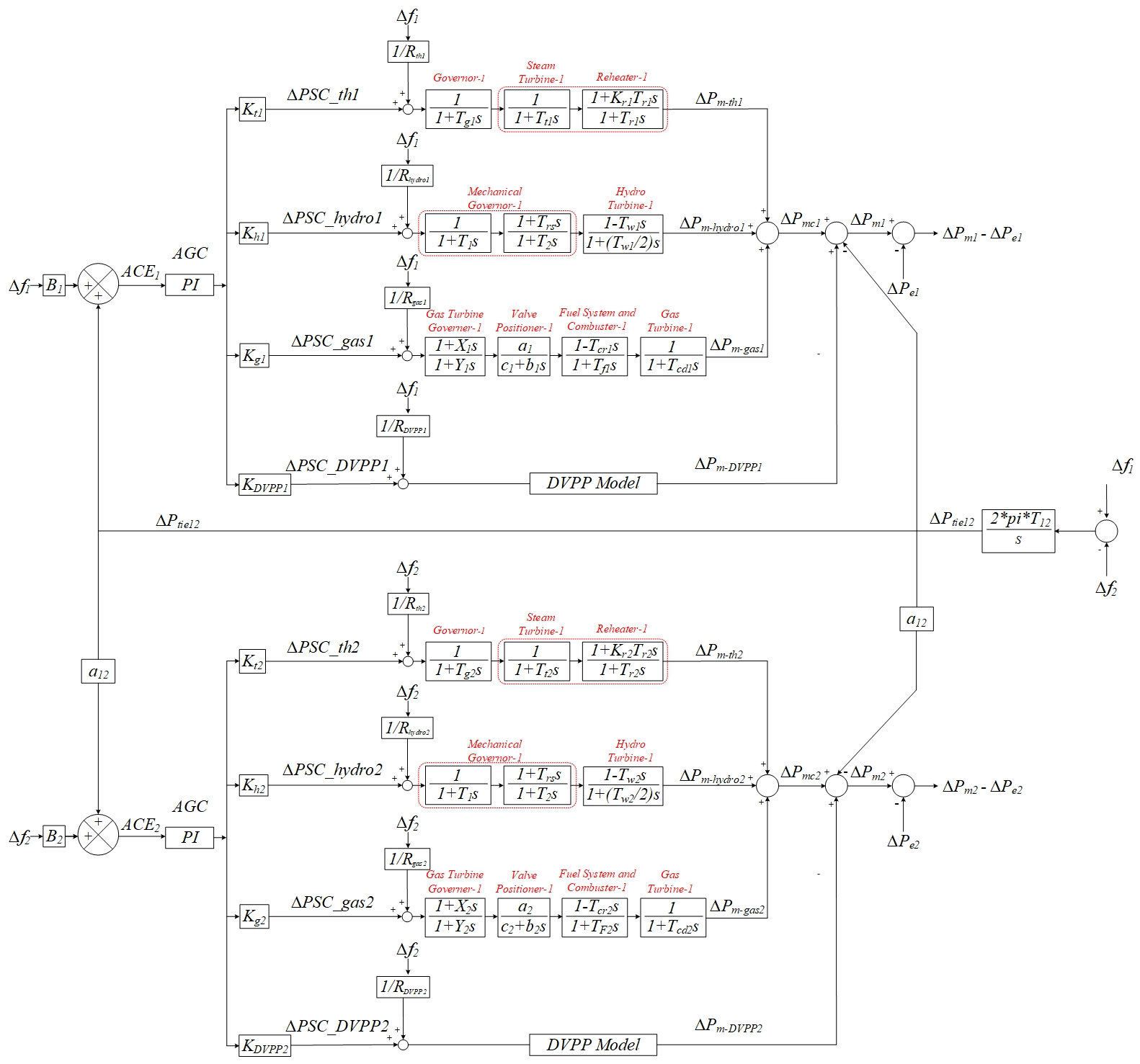}}
\caption{New control model for direct participation of the DVPP to a two areas SFC\label{figSFCtwoarea}}
\end{figure}

Fig. \ref{figDynamic_vf} shows the block diagram for grid dynamic equivalent implemented based on Eq. \ref{GDEswing} and Eq. \ref{GDEvoltage}. The difference between the global active power produced in the system and the global load of the system ($\Delta P_{m}$ - $\Delta P_{e}$) is considered as the input for the grid dynamic equivalent block. This input is achieved through the output of control architecture brought in Fig. \ref{figSFCtwoarea}.

Fig. \ref{figSFCtwoarea} shows the control model for generic two-area power system with direct participation of DVPP to SFC. This figure considers 2 zones but it can be extended to stand for any number of zones. The power system in each area contains the classic synchronous generators (thermal, hydro, and gas), and DVPP. The classic synchronous generators (thermal, hydro, and gas) are mathematically modelled by transfer functions as given in \cite{IndirectSFC1}, and \cite {IET}. In this paper, the two wind turbines type IV (PMSG based wind turbine model) are modeled in detail to take in to account the role of DVPP and all dynamic interactions. Notice that the modeling adopted here allows all the dynamics, including the grid voltage one, the DVPP generators primary ones and the rapid ones of the power electronics which connect the DVPP generators to the grid (Fig. 7) and not only the frequency dynamics as in previous SFC work in the literature (e.g., \cite {IndirectSFC1} , and \cite {IndirectSFC2}). Indeed, in our model, the full model of real PMSG machines have been taken in to account, and all the other primary controls for both frequency (including both fast control and very fast control (RoCoF)) and voltage are considered. Since we considered the real model of the machines, it is possible to consider the primary dynamics for voltage. On grid side, the model mentioned in Section \ref {SectionControlModels} provides mixed voltage and frequency dynamics.

As it can be observed from Fig. \ref{figSFCtwoarea}, DVPP directly participates to SFC, in the same manner as classic synchronous generators (the concept is shown in Fig. \ref{figDISFC}). The area control error of each zone will be regulated to zero by an effort shared among the classic synchronous generators and the DVPP's ones with participation factors of $K_{t}$,$K_{h}$,$K_{g}$,$K_{DVPP}$ as discussed in Section \ref{Dirct Participation}.

\section{Simulation results}

The simulation results are given in this section for direct participation of the DVPP to the SFC, tested on the benchmark represented in Fig \ref{figBenchmark} and implemented in Matlab/Simscape. It should be noted that a DVPP is only considered in one area (area 1) of Fig. \ref{figSFCtwoarea}. Therefore in the control model shown in Fig. \ref{figSFCtwoarea}, three classic equivalent synchronous generators (thermal, hydro, and gas)  and two detailed PMSG based wind turbines (representing DVPP, n=2), have been used in area 1. Grid frequency is provided by the grid dynamic equivalent block shown in Fig. \ref{figDynamic_vf}. Area 2 contains only three classic equivalent generators (thermal, hydro, and gas).

The inputs to SFC loop are the area control errors related to area 1 and area 2 (ACE1, ACE2).

DVPP participation is tested in two situations: \textit{nominal run} where all DVPP generators are normally (as scheduled) operating and \textit{disturbed situations} in which some generators of the DVPP are lost (due to, for example, lack of natural ressources - wind in our case - on some part of the DVPP) and internal DVPP redispatch is acting. In this section,  also some other disturbances, currently considered in power systems analysis like grid short-circuits, are also considered.

\subsection{Nominal scenarios}

As illustrated in the Fig. \ref {figSFCtwoarea}, the DVPP directly participates to SFC. In the nominal mode the load of the area 1 in the benchmark is 100 MW. The two PMSG-based wind turbines have rated power of 4.1 MW, each of them working in deloaded operation (at 3.5 MW), in order to provide reserve for frequency support.

\begin{figure}[!h]
    \centering
    \includegraphics[scale=0.3]{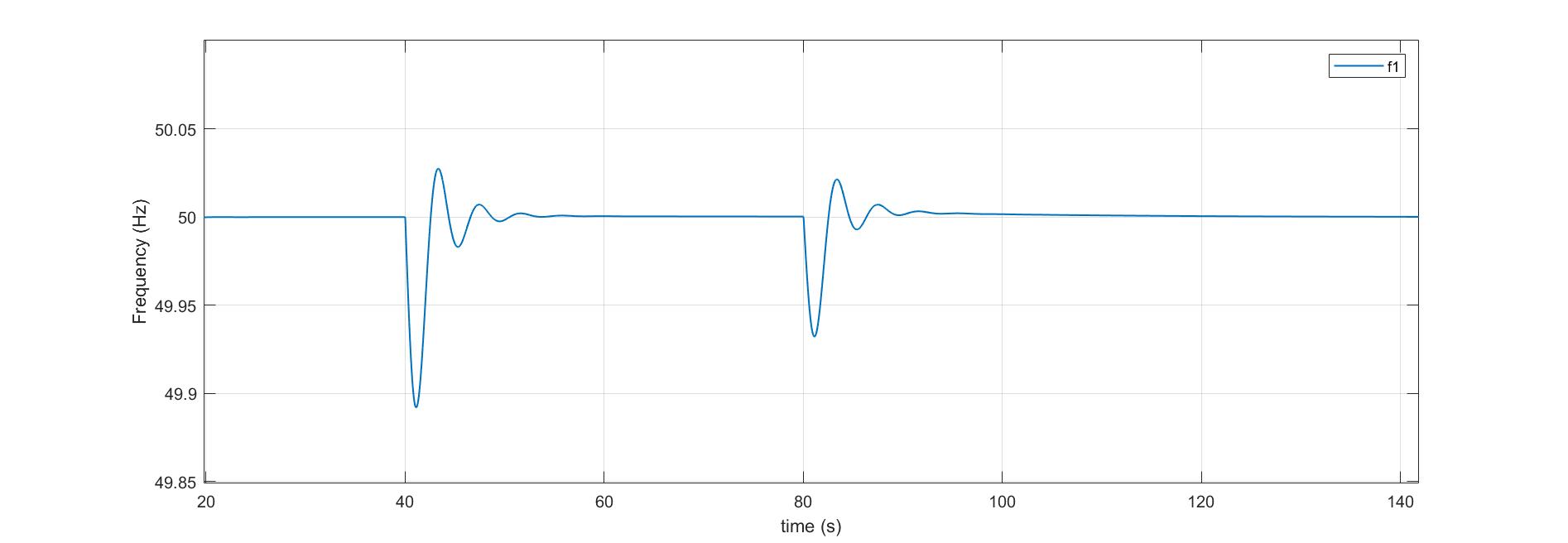}
    \caption{The Frequency of area 1}
    \label{figf1}
\end{figure}

An under frequency event has been considered as case study to test the SFC strategy. At t=40 sec, the active power of the load in the area 1 is increased by 6 percent (to 106 MW), causing the under frequency deviation shown in Fig. \ref{figf1}. 

It is assumed that 90 percent of frequency support will be provided by the classic synchronous generators, and 10 percent of frequency support will be provided by the DVPP. Therefore, the participation factors in Fig. \ref{figDISFC} are considered 0.1 for DVPP, and 0.3 for thermal, hydro and gas synchronous generation units.

\begin{figure}[!h]
    \centering
    \includegraphics[scale=0.3]{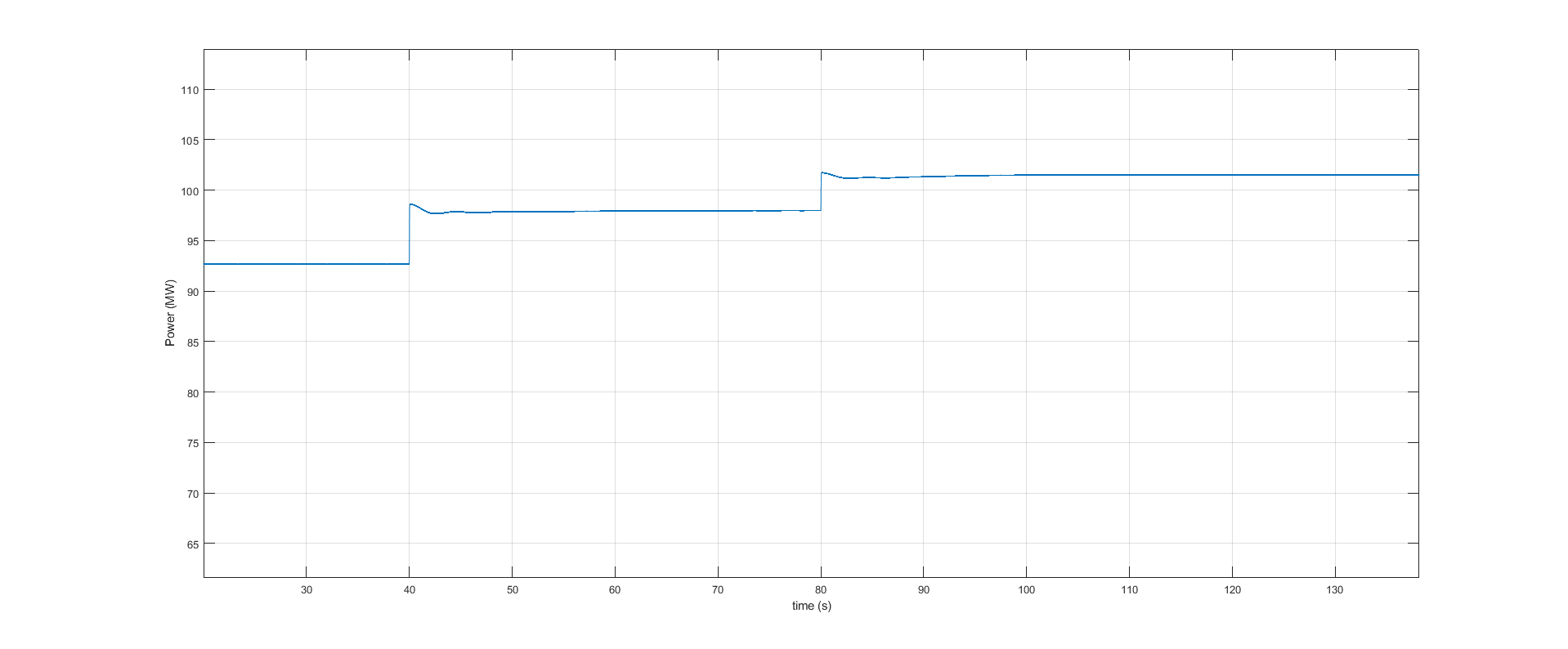}
    \caption{The active power generated by conventional synchronous generators units in the area 1}
    \label{figPE1}
\end{figure}

\begin{figure}[!h]
    \centering
    \includegraphics[scale=0.3]{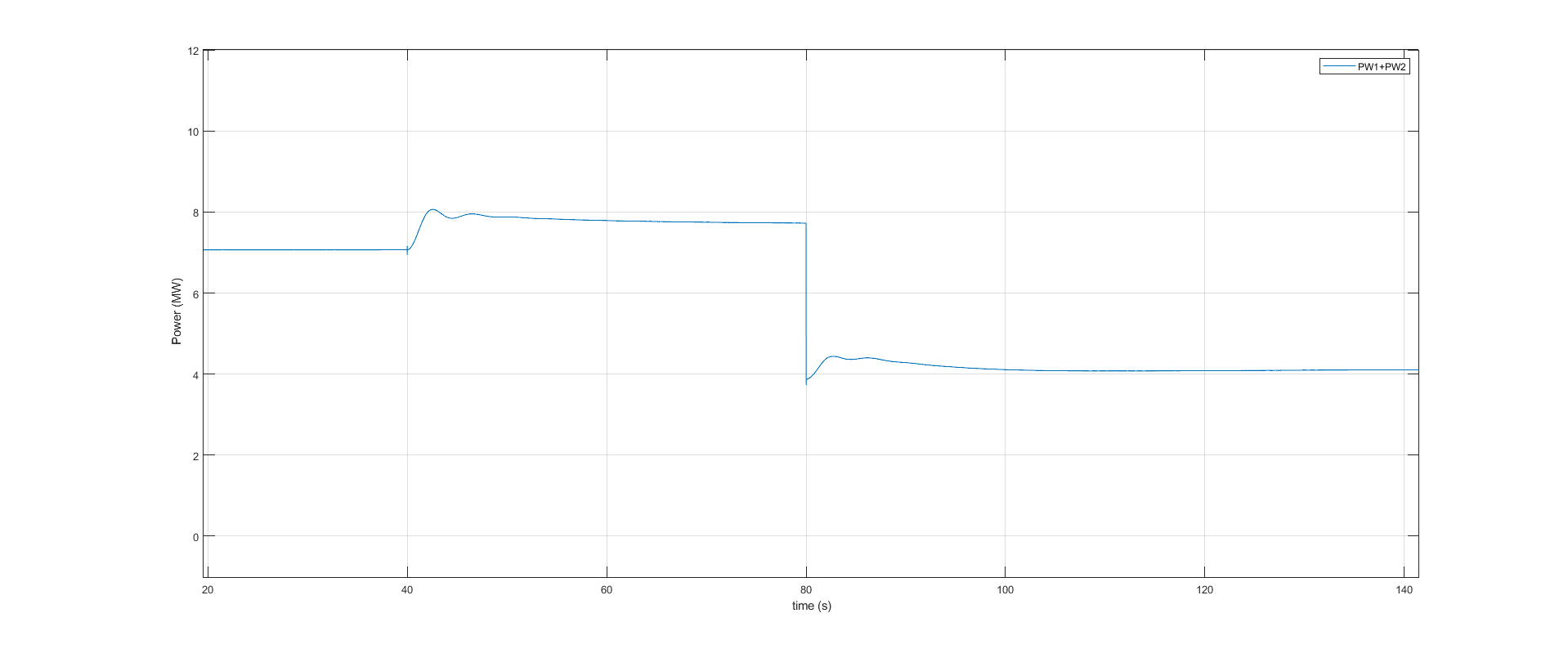}
    \caption{The active power generated by DVPP unit}
    \label{figPW12S}
\end{figure}

Fig. \ref{figPE1} and Fig. \ref{figPW12S} show the active power generated by the classic synchronous generators (thermal, hydro and gas) and the DVPP (PW1+PW2: the sum of active power generated by the two PMSG based wind turbines), respectively.

As it can be observed from Fig. \ref{figPE1} and Fig. \ref{figPW12S}, the classic synchronous generators and the DVPP will participate directly to SFC by increasing their power by 5.4 MW and 0.6 MW, respectively.

One can also notice nominal frequency response, compliant with the classic specifications of the SFC (Fig. 10).

The internal dispatching algorithm ensures that each WT will contribute to frequency support regarding its generating active power as given in Eq \ref{Redispatch}. Therefore, as it can be observed from Fig. \ref{figPW12}, both wind turbine will work at deloaded mode (PW1=PW2=3.5 MW) until t=40 sec. At t=40 sec, both wind turbines increase their power to provide the 10 percent of total active power required for SFC support (5 percent will be provided by each wind turbine). 

\begin{figure}[!h]
    \centering
    \includegraphics[scale=0.3]{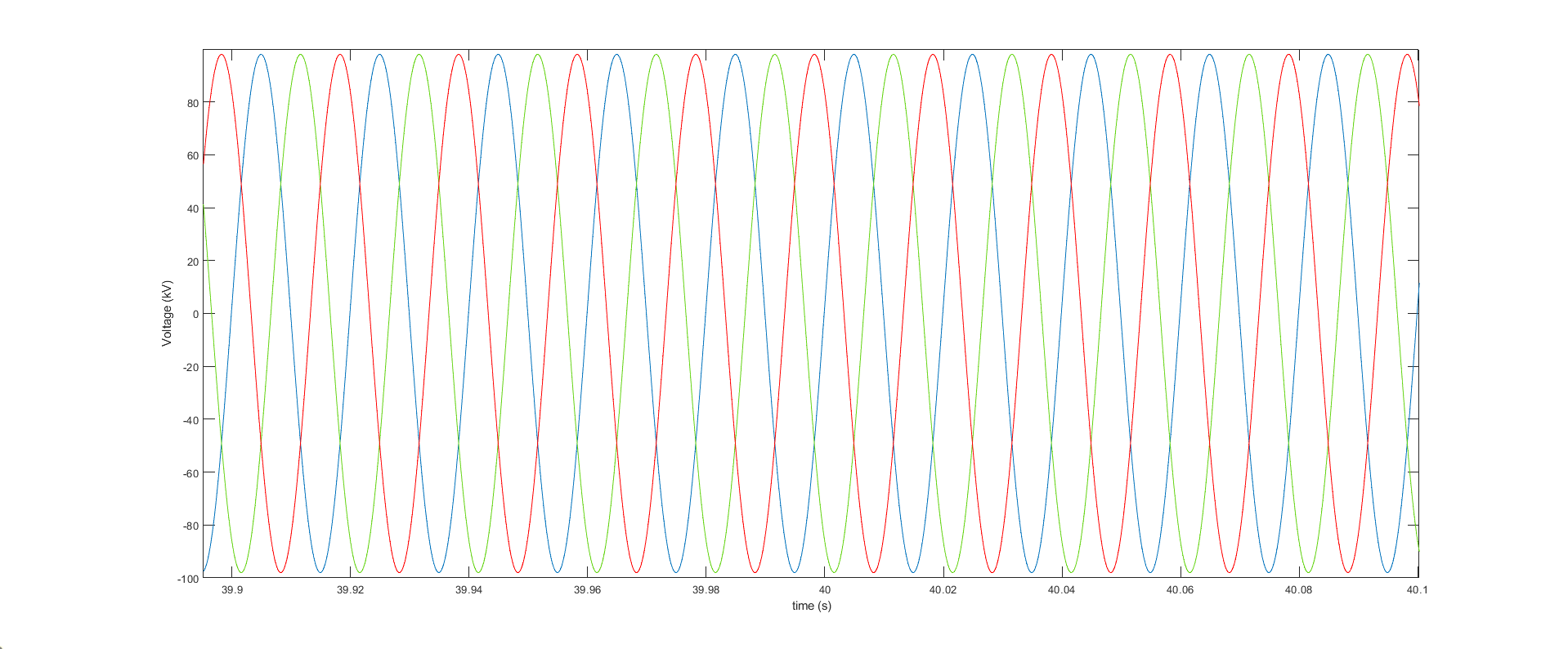}
    \caption{Three phase voltages at bus 3 of the benchmark}
    \label{figVbus3}
\end{figure}

Fig. \ref{figVbus3} shows the three phase voltages at bus 3 of the benchmark. It is worth mentioning that the amplitude of the voltage is constant at 1 per unit, while the frequency reduces due to the load increase, which can be observed from Fig. \ref{figf1}.

\subsection{Behavior in case of disturbances}

\subsubsection{Influence of internal redispatch}
\begin{figure}[!h]
    \centering
    \includegraphics[scale=0.3]{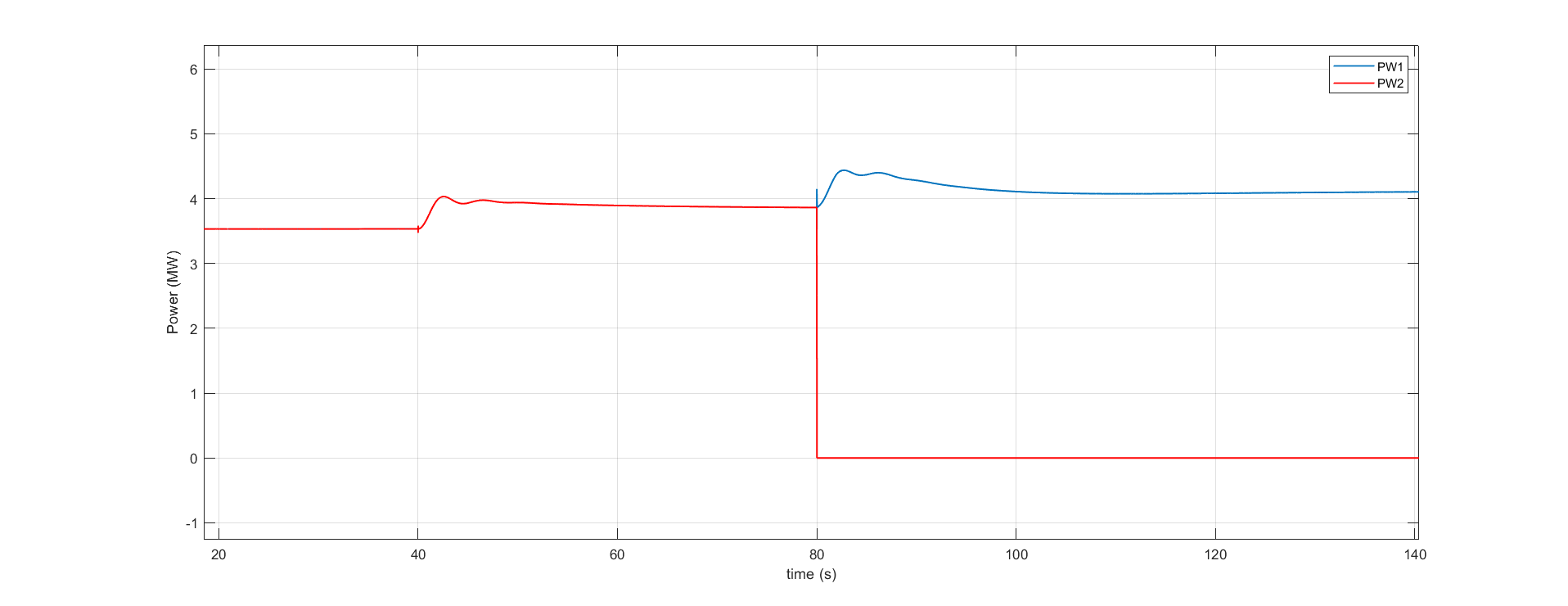}
    \caption{The active power of each PMSG-based WT}
    \label{figPW12}
\end{figure}

In order to test the  participation of the DVPP as a whole unit in provision of active power for secondary frequency support, at t=80 sec it is supposed that wind turbine number 2 goes out of service.  The internal redispatch regulation introduced in Section \ref{Internal Redispatch} will transfer to wind turbine number 1 all required SFC support (10 percent of total active power required for SFC support) (Fig. \ref{figPW12}). As it can be observed from Fig. \ref{figf2new}, at t= 80 sec, this internal redispatch has very little influence on the frequency dynamics.

It should be noted that the internal redispatch algorithm introduced in Section 3.2 works in the sense that it transfers the regulation effort to remaining WT1 in order to maintain the full participation of DVPP in secondary frequency support. However, in the case treated here the DVPP participation is limited by the predefined deloading margin of WT2. The latter increases its production in order to compensate the loss of WT1 but  the total the active power generated by DVPP decreases by 3.5 MW when WT2 goes out of service. This amount of active power is compensated by the classic generators via the AGC classic loop, as it can be observed from Fig. \ref{figPE1}.   

\subsubsection{Interaction between SFC zones}
\begin{figure}[!h]
    \centering
    \includegraphics[scale=0.3]{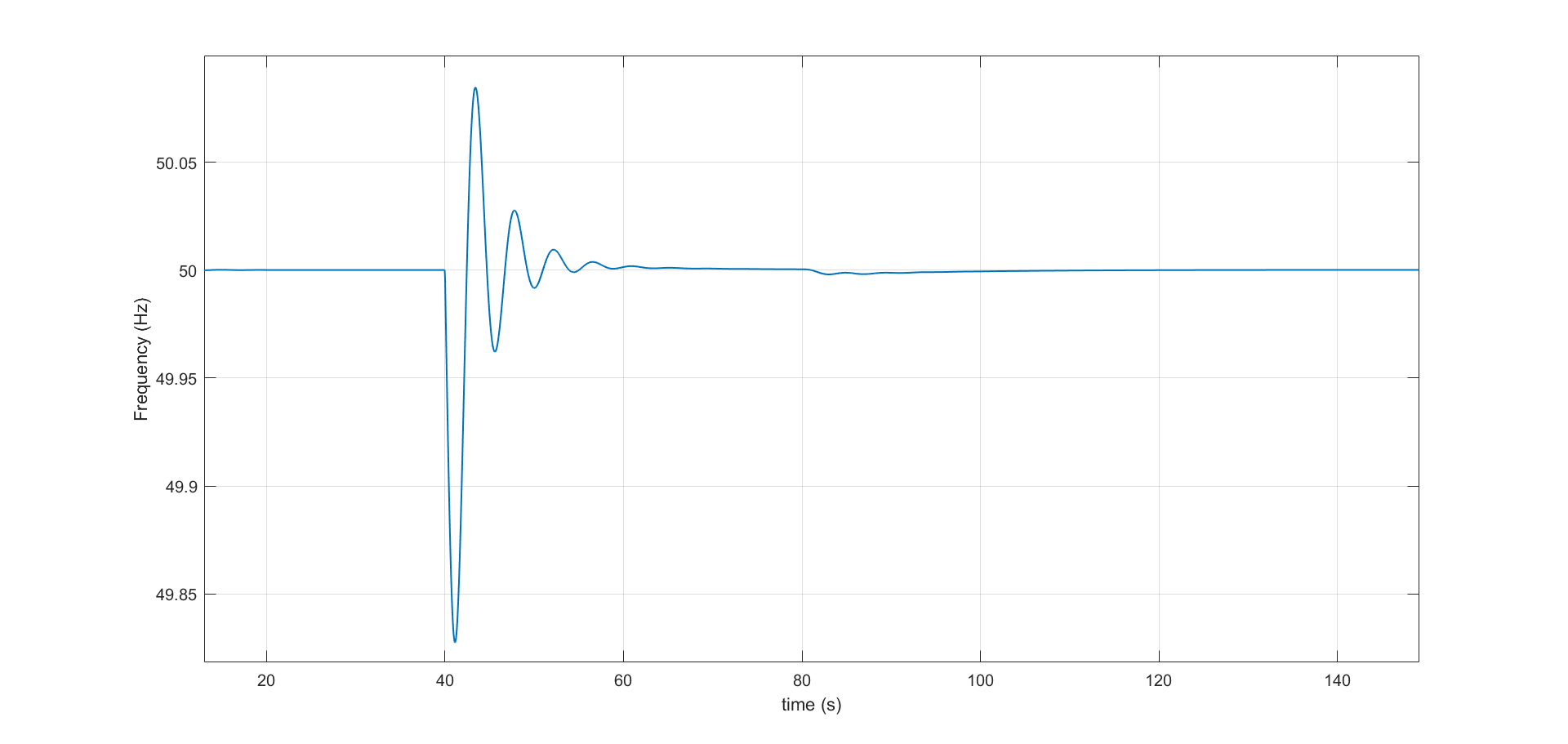}
    \caption{Frequency waveform in the area 2}
    \label{figf2new}
\end{figure}

In order to study dynamic interactions of the 2 areas controls, along with under frequency event in area 1, an under frequency event is also simulated (at t=40s) in  area 2 by increasing the active power of the load in this area by 10 percent. This is rejected in steady state in area 2 as shown in Fig . \ref {figf2new} (response to the event at t=40s). 

\begin{figure}[!h]
    \centering
    \includegraphics[scale=0.3]{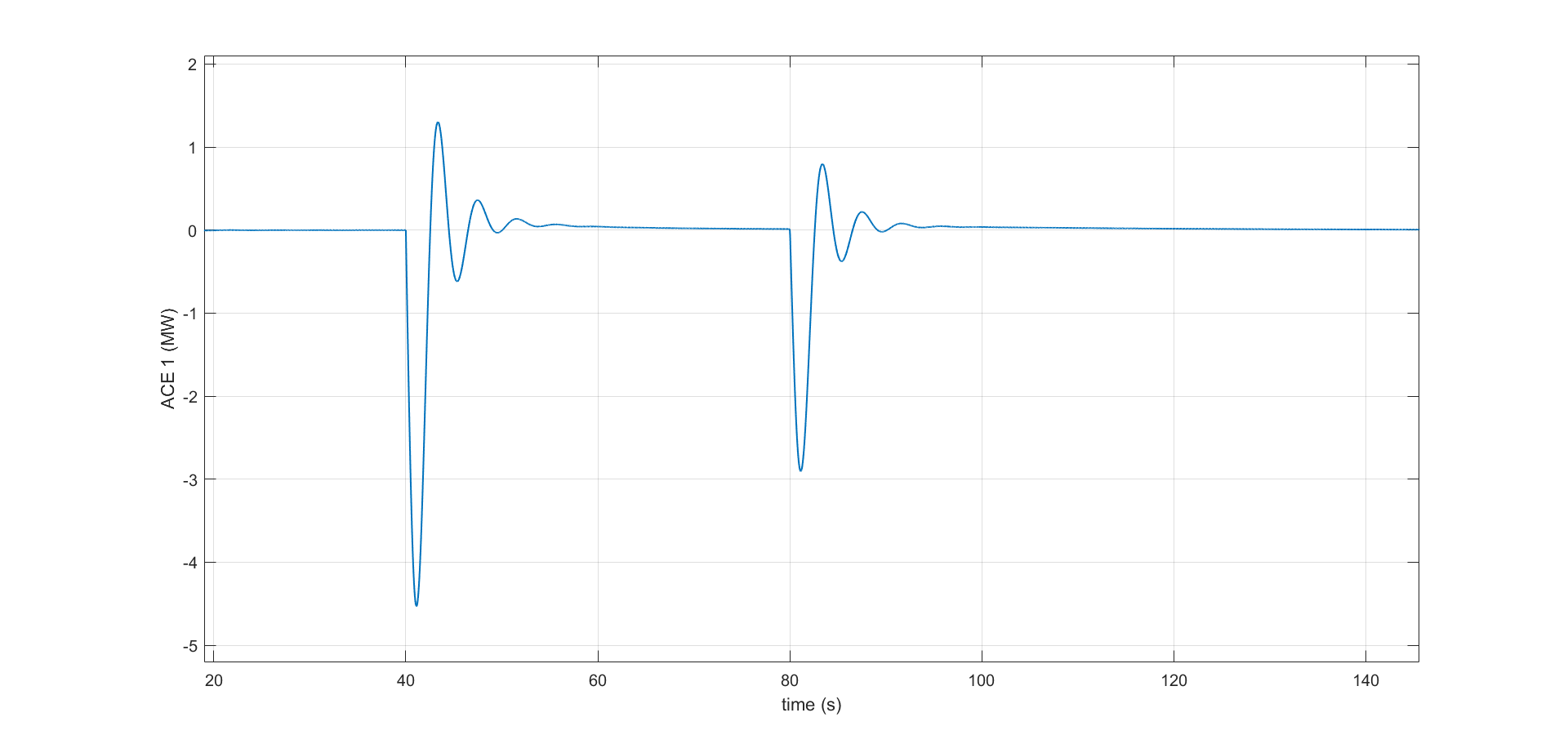}
    \caption{The area control error of area 1}
    \label{figACE1}
\end{figure}

\begin{figure}[!h]
    \centering
    \includegraphics[scale=0.3]{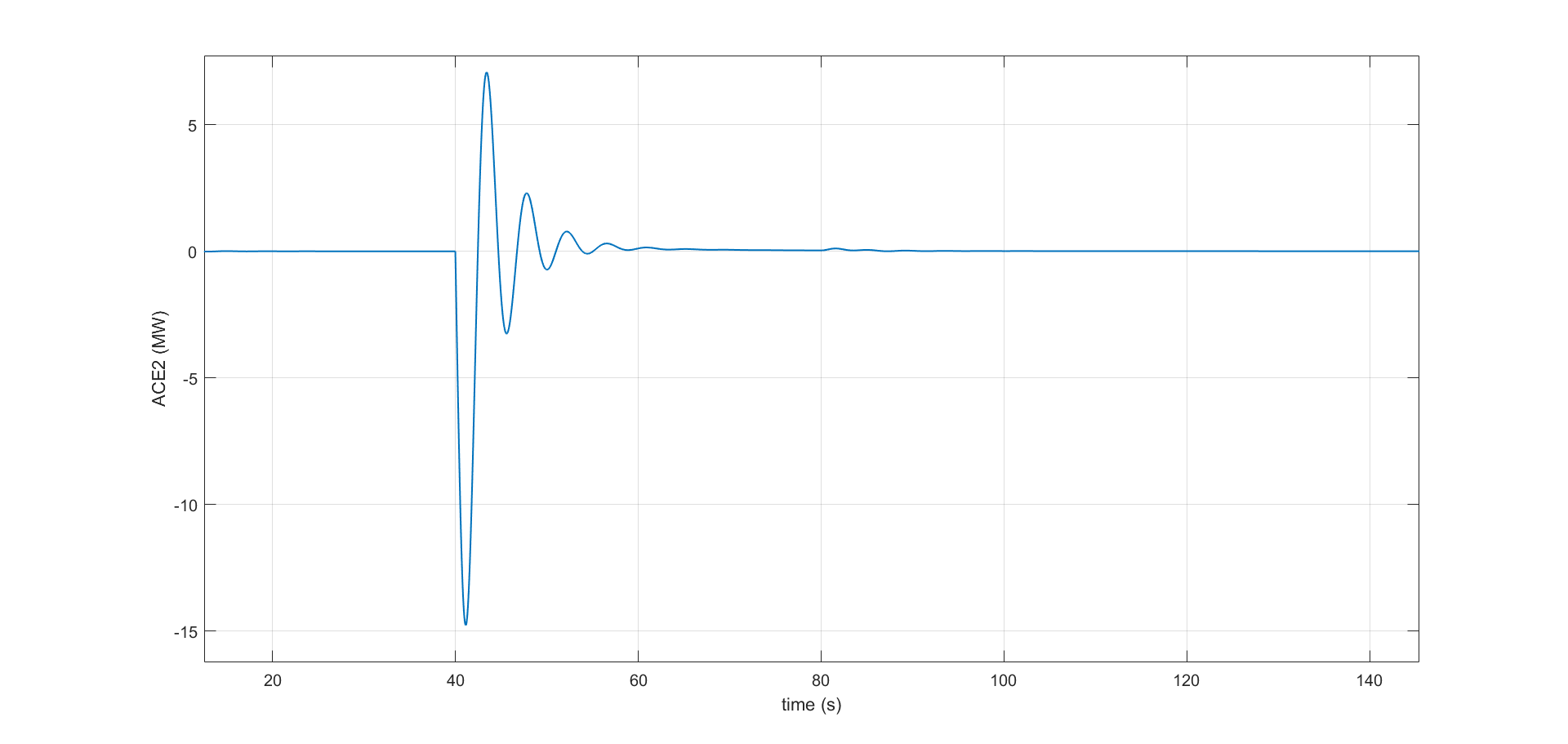}
    \caption{The area control error of area 2}
    \label{figACE2}
\end{figure}

Fig . \ref {figACE1} and Fig . \ref {figACE2} give the ACEs related to area 1 and area 2, respectively. As it can be observed from this figures, our proposed strategy efficiently eliminates the area control error in both areas.

\begin{figure}[!h]
    \centering
    \includegraphics[scale=0.3]{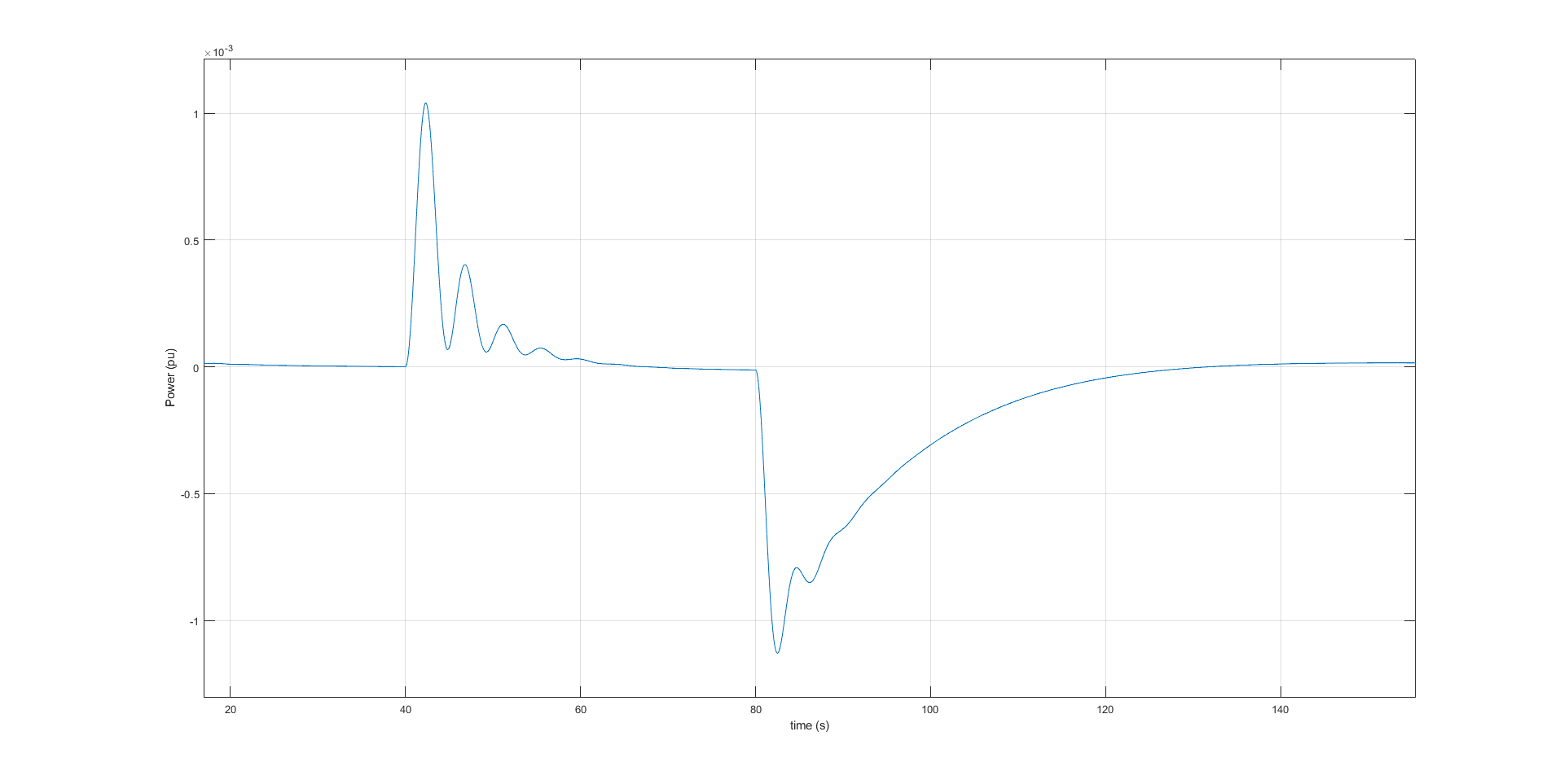}
    \caption{The tie-line power between two areas}
    \label{figDeltaptie}
\end{figure}

Fig . \ref {figDeltaptie} shows the tie-line power between the two areas. The reference for this is tracked without steady state error for all studied scenarios, including disturbances.

\subsubsection{Short-circuit test}
A short-circuit test is considered here to analyse the control response in case of large disturbances which excite also faster - voltage and frequency primary -  dynamics. Filters introduced in Fig. \ref{figDispatching} are intended to avoid bad interactions with such faster control loops.  

\begin{figure}[!h]
    \centering
    \includegraphics[scale=0.3]{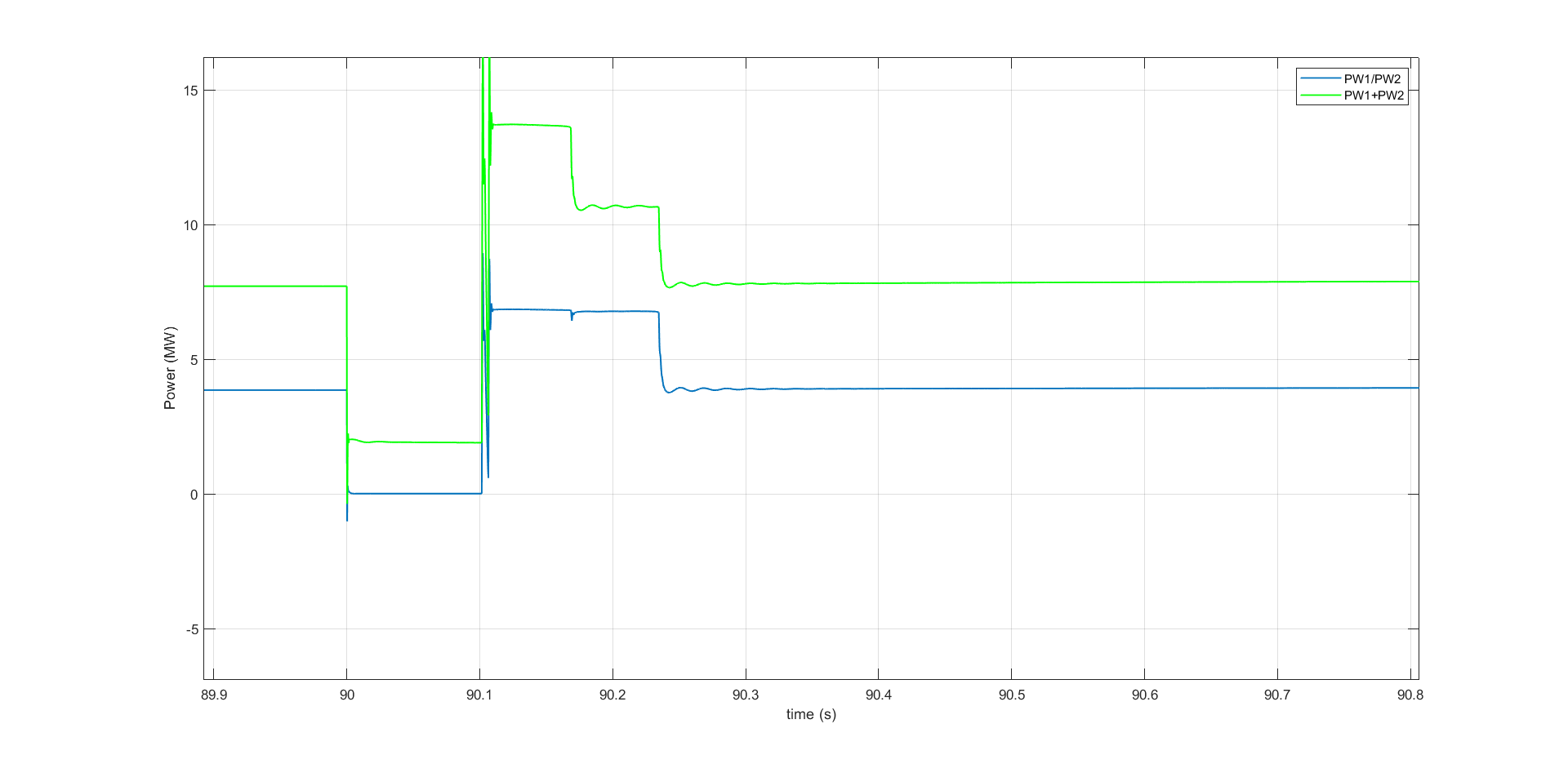}
    \caption{Active power of WTs and their sum in response to the three phase short-circuit}
    \label{figPW12SSC}
\end{figure}

Fig . \ref {figPW12SSC} shows the active power of WTs and their sum, when a 100ms duration three phase short circuit occurs at t=90s at the bus 1, which is the grid connection point of WT1 (Fig.\ref{figBenchmark}). As it can be observed from this figure, the active power smoothly returns to its steady state value after the short-circuit fault is cleared.

\begin{figure}[!h]
    \centering
    \includegraphics[scale=0.3]{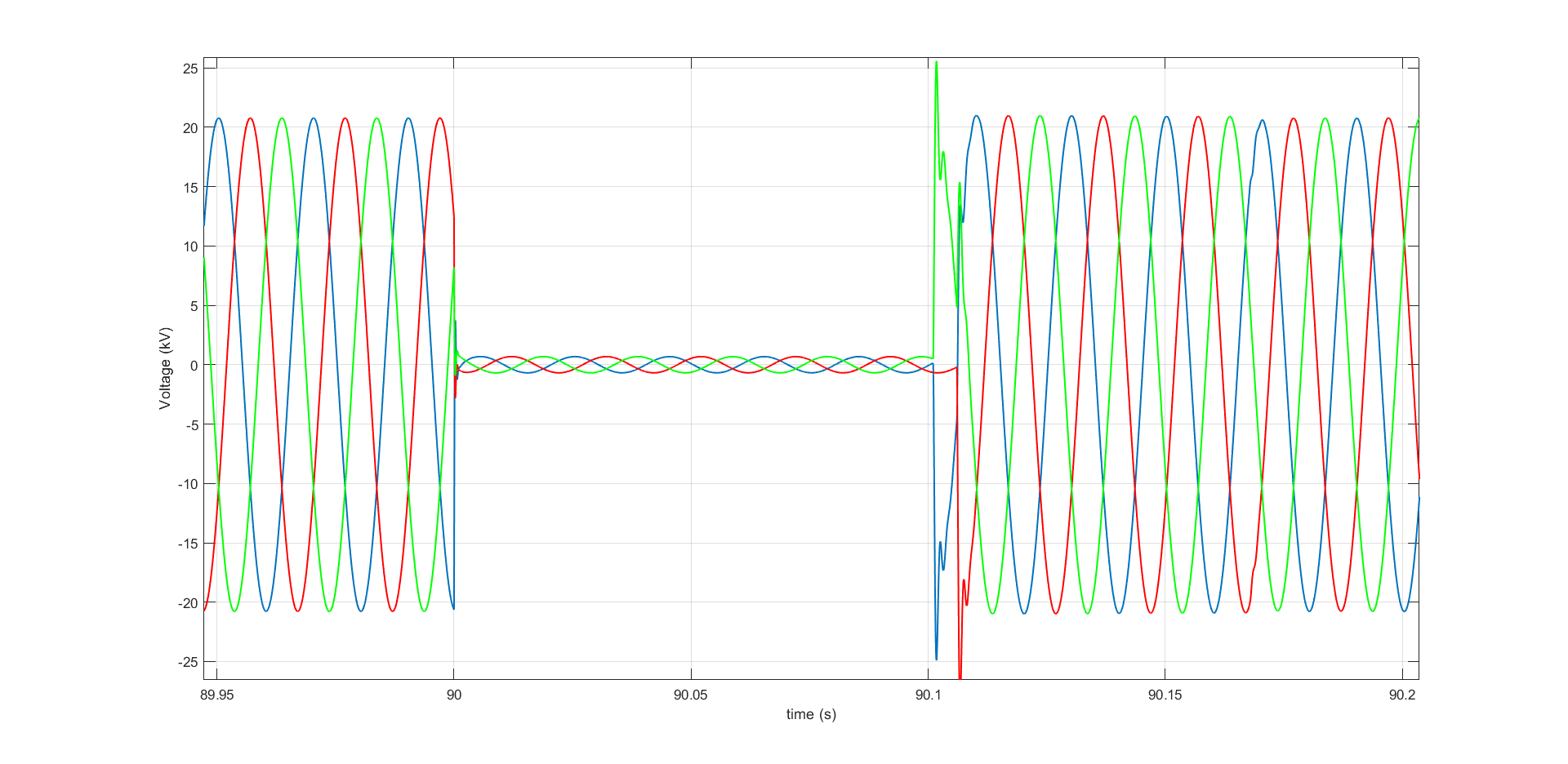}
    \caption{Voltage at bus 1 in response to the three phase short-circuit}
    \label{figVbus1SC}
\end{figure}

Moreover, Fig. \ref{figVbus1SC} shows the response of the rms value of the voltage of bus 1 to the considered short-circuit. One can notice no interaction with the primary voltage and frequency dynamics. This is due to the design of the internal redispatch filter (1/1+4s) shown in Fig. \ref{figDispatching}. More specifically, the time constant of this filter is chosen to well separate in frequency band the SFC response from the ones of the primary control loops.
Notice that, as explained in Section \ref{SectionControlModels}, a new model to represent mixed voltage and frequency dynamics is used in comparison to past approaches and studies for the SFC. This allows us to ensure the lack of interactions with rapid dynamics of power electronics and fast loops of classic synchronous generators at both synthesis and validation levels.

\begin{figure}[!h]
    \centering
    \includegraphics[scale=0.3]{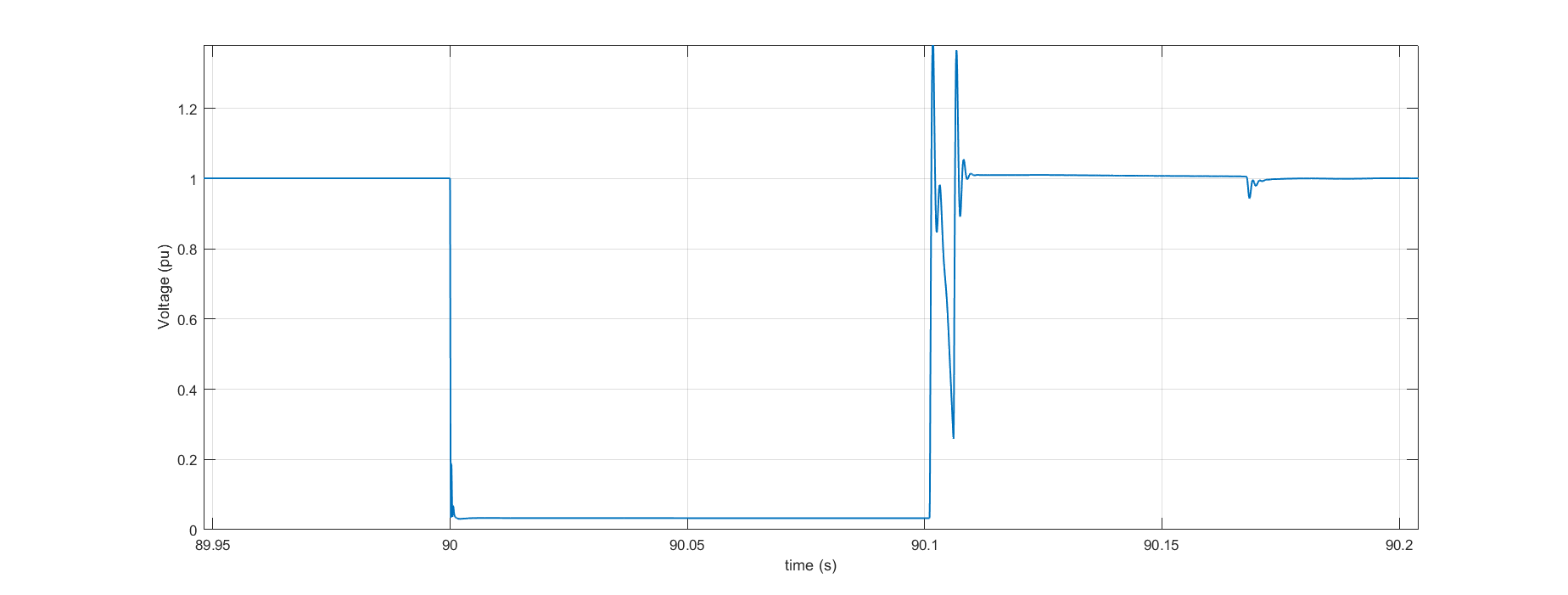}
    \caption{rms value of voltage at bus 1 in response to the three phase short-circuit}
    \label{figVbus1pu}
\end{figure}

\section{Conclusions}

In this paper, the direct participation of DVPPs for provision of secondary frequency support have been studied. A new SFC control strategy is proposed in which DVPPs are fully participating  in the same manner as classic synchronous generators. 

Also, an internal redispatching control algorithm is proposed to deal with inherent intermittance due to natural ressources (sun or wind) variations and accidental failures inside the DVPP. In such a way one could ensure reliable full participation to SFC of the whole DVPP as a single equivalent unit from both points of view of dynamic control and regulatory/market participation to frequency grid service.

The whole control is proven to be also robust against regular disturbances of a power system (like short-circuits) and not to interact with more rapid dynamics of power electronics via which the renewable generators (of the DVPP or outside) are connected to the grid or with neighbour classic synchronous generators (involved or not into the SFC).

Simulation results have been done on a two area power system to prove the effective performance of proposed SFC control strategy in provision of frequency support. These simulations are realistic from several points of view
\begin{itemize}
\item they consider a detailed model of the DVPP generators (PMSG machines in the example treated here)
\item the grid dynamics model integrate both voltage and frequency dynamics
\item two SFC zones are considered to incorporate the tie-lines power exchange regulation and to consider dynamic interactions at all levels (frequency/time-domains) between the SFC zones.
\end{itemize}

All the latter points are original in integration of renewables in the SFC.

Future extensions of this work will be done to consider more detailled DVPPs (with more than 2 generators and specific topologies (e.g., with RES generators connected to distribution grids also)) as well as hardware in the loop validations of the resulting control.




\section{Acknowledgements}
  This project has received funding from the European Union’s Horizon 2020 research and innovation programme under grant agreement No. 883985 (POSYTYF -- POwering SYstem ﬂexibiliTY in the Future through RES, https://posytyf-h2020.eu/).
\end{document}